\documentclass[11pt,twoside]{article}
\usepackage{asp2010}

\resetcounters

\begin{document}

\title{Extended atmospheres of AGB stars: modeling and measurement}
\author{Michael J. Ireland$^1$}
\affil{$^1$Sydney Institute for Astronomy (SIfA), School of Physics,
  A28, University of Sydney, NSW 2006, Australia}

\begin{abstract}
Encoded in the time- and wavelength dependent properties of pulsating
AGB stars are the underlying fundamental parameters of mass,
composition and evolutionary state. However, the standard technique of
placing stars on a HR diagram, even with the aid of pulsation periods,
can not be done easily for extended AGB stars, because of the
difficulty of defining a radius or temperature. The atmospheres of
Mira variables are so extended that the optical depth unity radius can vary
by a factor of ~3 over the energetically important region of the
spectrum. Many important constituents in the radiative transfer are
far from local thermodynamic equilibrium, and for the coolest stars,
the process of dust formation and destruction requires a
time-dependent model of grain growth. I will describe the challenges
and some of the solutions to modeling these atmospheres, and describe
the utility of different kinds of observations in helping understand
both fundamental parameters and chaotic processes in complex AGB
atmospheres. 
\end{abstract}

\section{Introduction and Scope}

The Asymptotic Giant Branch (AGB) is the evolutionary state where the
majority of stars ending their lives today lose the majority of their
mass. It is primarily for this reason that they naturally have extended
atmospheres, as a photosphere gradually transitions to a stellar wind. AGB
stars have much more extended atmospheres than mass-losing hot stars,
because winds can only become radiatively driven around AGB stars
after dust forms. In turn, this means that an additional mechanism,
{\em pulsation}, is required in order to elevate the material high
enough above the photosphere. The resulting picture is rather complex,
depending on chemistry, luminosity, evolutionary state, mass, and
including effects of time-dependent highly non-linear pulsation, highly
supersonic shocks and non-equilibrium dust formation.

It is due to this complexity that the scope of this review must be limited. I will
focus on Mira-type variables as the AGB stars with the most extended
atmospheres, and will not focus on the super-wind phase. I will spend
little time on C-stars both due to personal biases and a lack of as
much observational information, and will only consider spherically
symmetric properties and models. I
will discuss the unique measurements that are possible for pulsating
AGB stars in Section~\ref{sectMeasurement}, , I will discuss the state of modeling efforts and
challenges in Section~\ref{sectModelling}, and I will discuss the most
important and exciting future goals in Section~\ref{sectFuture}.

\section[Measurement]{Unique Measurements}
\label{sectMeasurement}

AGB stars have absolute K-magnitudes in the range -6 to brighter than
-10, and are the most numerous stars in this magnitude range for all
but the very youngest stellar populations. This means that they can be
seen to great distances. Their nonlinear pulsation and large angular size means
that they can in principle be studied in more detail than arguably any
single star other than the sun. In the following sections I will
describe a subset of reasons why observations of extended AGB
atmospheres are so unique.

\subsection{Nearby Details}
\label{sectNearby}

Within about 500\,pc, we can learn a lot about pulsating AGB stars
primarily because they can be resolved both spatially and spectrally
throughout the energetically important region of their spectra. The
smallest Mira variables, with diameters of $\sim$1\,AU subtend an
angular size of 2\,mas at 500\,pc distances, clearly resolvable on
200\,m baselines by both northern and southern hemisphere infrared
interferometers.

The early speckle interferometry measurements of \citet{Labeyrie77}
pioneered the study of resolved extended AGB atmospheres, showing that
the apparent angular size of the eponymous Mira $o$~Ceti varied by a
factor of $\sim$2--3 in and out of TiO absorption features. As will be
discussed below, this rather incredible result, when combined with
spectra, only makes sense if the dust that is thought to drive the
stellar wind is optically thin at all wavelengths, and if the strong
TiO features are far from local thermodynamic equilibrium.

Subsequent years have produced many more interferometric results than
can be effectively modeled, including: aperture-masking
interferometry observations of 10 nearby Miras at $<$1\,$\mu$m
wavelengths by \citet{Haniff95}, diameter measurements of 18 nearby
Miras at 2.2\,$\mu$m by \citet{vanBelle96}, detailed 
wavelength-dependent
\citep{Mennesson02,MillanGabet05,Ireland05,Woodruff09} and
time-dependent diameters  
\citep[e.g.][]{Woodruff08}, and the first ``true'' images of a
Mira where the continuum photosphere and shell are well resolved 
(looking rather spherical after all that hard work) by
\citet{leBouquin09}. The common observational
theme in these measurements is that most, and sometimes all, wavelengths,
have a center-to-limb variation with more than one significant
component, and that the wavelength-dependence of diameter is more
significant than phase- or cycle-dependence. This
wavelength-dependence of diameter means that a radius is difficult to
define, and consequently, and effective temperature is {\em not
  meaningful} in the same way as for compact stars.

\begin{figure}
\plotone{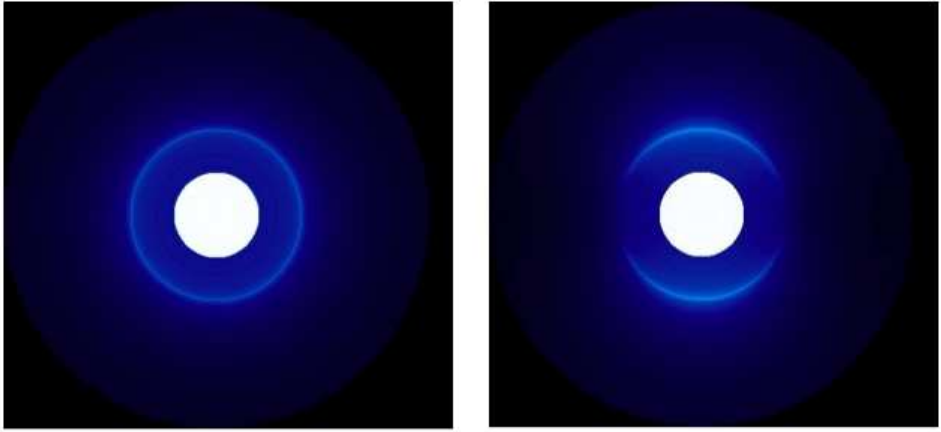}
\caption{Left: A toy model of a star with dust shell and
  radiatively-driven wind as seen in unpolarized light. Right: the
  same model as seen in polarized light with the E-field oriented
  horizontally.  Despite this symmetrical
  model having zero total polarization, the brightness of the shell as
resolved by an interferometer is different depending on if the baseline
is parallel or perpendicular to the baseline.}
\label{figPol}
\end{figure}

A relatively new use of optical interferometry for nearby Miras is to
make optical interferometry polarimetry observations, in order to
separate scattered light from thermally emitted light. An illustration
of how this is possible is given in Figure~\ref{figPol}. These
observations are particularly important because radiation pressure on
dust is thought to be the main driver of AGB stellar winds, and
observing the light scattered by dust determines the dust properties
and distribution.

\begin{figure}
\begin{center}
\plotone{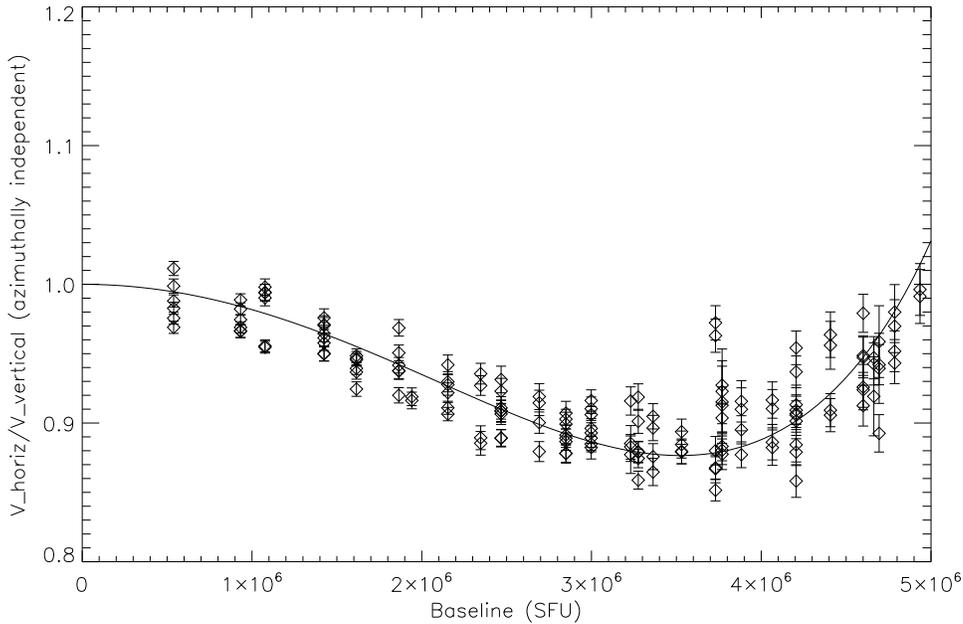}
\end{center}
\caption{The visibility ratio for polarizations parallel and
  perpendicular to the interferometric baseline for W~Hya at a
  wavelength of 1.24\,$\mu$m (Norris et al 2011, in
  preparation). The fit (solid line) is a thin shell model with 
  $\tau_{\rm scat} \sim$0.1 and a shell radius of $\sim$2 stellar
  radii.}
\label{figNorris}
\end{figure}

This kind
of observation was pioneered by \citet{Ireland05}, who demonstrated
that dust forms only $\sim$2 stellar radii from two Mira variables,
R~Car and RR~Sco,
and scatters $\sim$20\% of the stellar flux at 900\,nm. The SAMPol
mode on the CONICA camera at the VLT has extended the availability of
this method to the near-infrared at significantly higher
signal-to-noise (Norris et al. 2011, in preparation), as shown in
Figure~\ref{figNorris}. A preliminary result from these
multi-wavelength observations
is that there is a population of grains at $\sim$2 stellar radii that
have grain radii of order 300\,nm. 

In addition to optical and near-infrared interferometry, there are
many other unique observations available of nearby Mira
variables. Mira variables are some of the only nearby stars that have
their continuum photospheres resolvable in the radio
\citep{Reid97}. These kinds of observations are much easier now with
upgrades such as EVLA in the northern hemisphere and CABB in the
south. These observations probe the ionization edge of metals such as
Na and K in the extended photosphere. Masers in turn probe the
velocity structure of the atmosphere \citep[e.g.][]{Cotton08}.

By virtue of their brightness and appeal to amateur astronomers,
nearby Miras also have long photometric time-baselines, up to hundreds
of years. This means
that period are precise despite the long periods, and long-term period
changes (possibly indicating recent thermal pulses) are detectable
\citep[e.g][]{Zijlstra02}. 

Finally, high spectral resolution observations
\citep[e.g.][]{Hinkle97} have been under-utilized, mainly
because of the difficulty in interpretation, and have only been made
in select wavelength intervals. These are of course relatively easy
observations to make even near minimum-light in the visible, and
contain a wealth of information due to the various spectral features
that probe different heights in the atmosphere.

\subsection{Distant Complexity}

For distant Mira variables, particularly those outside the Galaxy, the
number of possible observations are much more manageable. Firstly,
pulsating AGB stars have all the measurements available for
non-pulsating stars, but can be seen to larger distances due to their
high luminosities \citep[e.g. Cen~A, ][]{Rejkuba04}. This will become
even more important with new space missions available for sensitive,
deep photometry, especially the JWST (the Virgo cluster, 
see Section~\ref{sectFuture}). The most important
spectral features are broad-band colors, continuum shapes and
molecular band strengths - these can all be determined at low spectral
resolution so are even easier to observe at large distances than stars
of earlier spectral type.

The most important additional observation available for AGB stars is
pulsation period. In principle, the combination of the four observationally
determined variables of composition (i.e. relative strength of
spectral features), temperature, apparent luminosity
and period is enough to uniquely determine four fundamental
parameters, e.g. composition, distance, luminosity and mass. Of
course, correct interpretation is dependent on identifying the correct
pulsation mode, and having trustworthy models! In addition, AGB stars
may have more hidden variables, such as core mass (which does not
uniquely determine luminosity in the presence of thermal pulses) and
redenning law. 

Going one step further, the large amplitude pulsation that produces
the extended AGB atmospheres means that nonlinear phenomena are
constrained by observations of pulsation amplitude and light-curve
shape. These observations are necessarily made whenever a period is
measured. This information has been used in the case of Cepheid variables
in the Large Magellenic Cloud (LMC) by \citet{Keller06} to uniquely
constrain all fundamental parameters including the distance to the
LMC. For the case of Mira variables, however, period-luminosity-color
relationships have been used successfully \citep{Feast89}, but
period-luminosity-color-amplitude or more complex relationships have
not. Calibration of such relationships and a quantitative
understanding of the width of the Mira period-luminosity relationship are
needed before these observations of distant AGB stars can be used optimally.

\section[Modeling]{Modeling Challenges}
\label{sectModelling}

Making comprehensive models of pulsating AGB stars is the only way to
understand underlying physical parameters of initial mass, composition
and evolutionary state. There is some hope to
calibrate models indirectly, e.g. statistical initial masses of a
population using kinematics \citet[e.g.][]{Wyatt83}, initial
composition of cluster Miras or confirming that the same models work
for low-amplitude cluster Miras \citep[e.g.][]{Lebzelter07} or
possibly smaller-amplitude AGB variables in binary stars. Masses of
individual Mira variables are not likely to be measured from binary
orbits due to the very long periods involved. Even for Mira itself, a
relatively wide binary with a $\sim$1000 year period, wind accretion
phenomena onto the secondary and pulsation in the primary 
would complicate obtaining a spectroscopic orbit even if the time baseline
were available.

\begin{figure}[hb]
\plotone{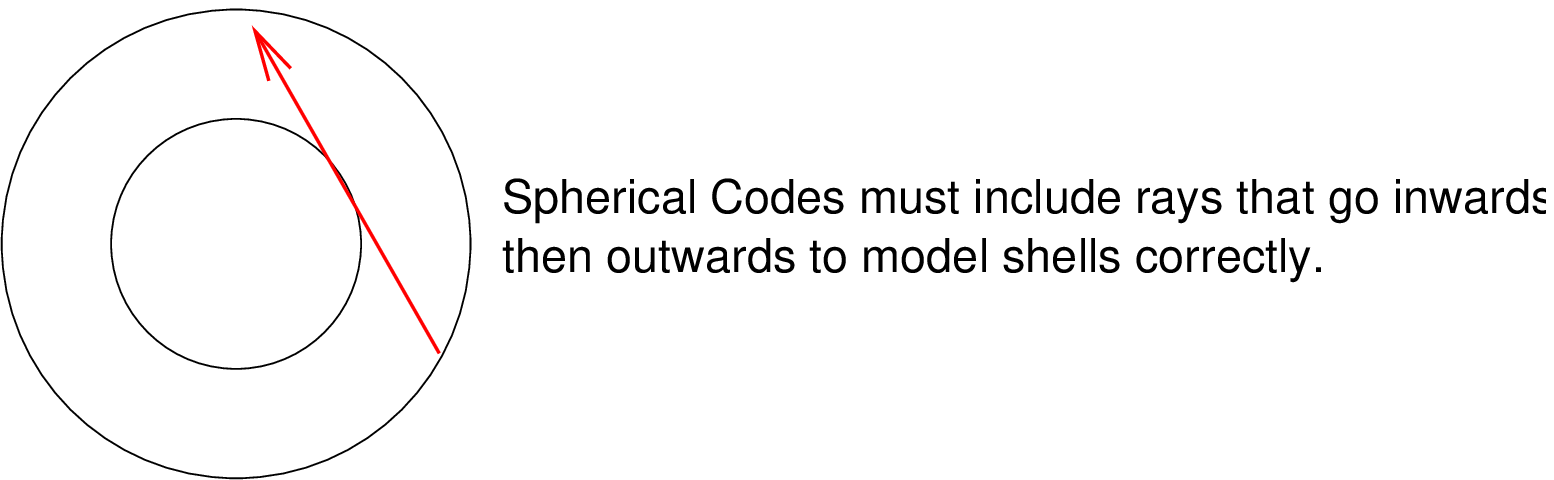}
\caption{The kind of ray that is key to solving for radiative
  equilibrium in an extended spherical code. For purposes of
  computational efficiency, different spherical codes use different
  number of these rays or other approximations, and may not be suited
  to the very extended configurations in Mira variables.}
\label{figSpherical}
\end{figure}

Specialized model atmospheres are needed for Mira variables because
the atmospheric extension is so important in interpreting spectra and
other observations. For
example, water shells are observed at radii of $\sim$2 times the
continuum photospheric radius \citep[e.g.][]{leBouquin09}. This means 
that the
geometrical dilution of the radiation field at the location of this shell gives a
factor of $\sim$4 times less flux as compared to an equivalent shell
located just above the continuum photosphere, and means that the
temperature of this layer is $\sim$30\% cooler than it would be in a
compact model. This extension can not be produced by a static model,
so the run of density with radius must come from a dynamical pulsation
code. The radiative transfer calculation therefore involves a
spherical atmosphere that is poorly approximated by the plane-parallel
case (e.g. Figure~\ref{figSpherical}).
Although many static model codes \citep[e.g. MARCS,
][]{Gustafsson08} are spherical, they have not generally been used to
model atmospheric structures as extreme as Mira variables, and do not
have an easy way to interface with dynamical codes.

Several groups have produced competitive dynamical, extended
atmosphere codes for modeling AGB stars. Codes that are currently in
use are the Cool Opacity-sampling Dynamic EXtended ({\tt CODEX})
models by the Sydney-Canberra-Heidelberg group \citep{Ireland08}, and
the Uppsala code \citep[e.g.][]{Hofner03}. The {\tt CODEX} models have
only been used to model M-type Miras, while the Uppsala code has been
mostly used for C-type Miras, but has more recently been applied to S-
an M-type Miras also. \citet{Woitke06} and \citet{Jeong03} described codes
that had relatively simple radiative transfer but sophisticated O-rich
grain growth descriptions. To the author's knowledge, these codes are
not in active use today.

Some of the key differences between the two codes in active use are given in
Table~\ref{tabCodes}. At the time of writing, there are various
compromises made in each case. The {\tt CODEX} models models are a
marriage between detailed gray self-excited pulsation models
\citep[e.g.][]{Keller06} and an opacity-sampling code that re-solves
for temperature in the outer atmosphere at selected model phases,
based on the method of \citet{SchmidBurgk84}. This
outer atmosphere code uses an opacity sampling method and dust formation approximations
only expected to be valid within a few continuum-forming radii, where
grain grown time-scales are small compared to the pulsation time-scale
\citep{Ireland06}. The Uppsala models simultaneously combine dynamics and radiative
transfer into a single large implicit difference equation scheme which
is solved with Newton-Raphson iteration. These models
includes the time-dependent growth of a single grain species, but has the capability
for much more sophisticated heterogeneous grain growth in the code
(H\"ofner 2010, personal communication). The Uppsala code introduces
pulsation with a sinusoidal piston at the base of the atmosphere,
which means that detailed phase-dependent results may not be reliable
and that it is more difficult to relate model properties to
fundamental parameters. Although the {\tt CODEX} models are
self-excited, there are two key free parameters in the models that have
not been calibrated, so their ability to relate model properties like
pulsation amplitude to underlying physical parameters is an
in-principle strength only.

\begin{table}
\caption{A comparison of key model properties of the two actively used
 dynamical extended atmosphere codes.}
\begin{tabular}{lll}
\hline
Model Property & {\tt CODEX} & Uppsala \\
\hline
Pulsation             & Self-Excited                   & Piston \\
Pressure Structure    & From Gray Model                & Self-Consistent \\
Temperature Structure & Opacity-Sampling               & Opacity-Sampling \\
                      & (4300 wavelengths)             & (64 to 300 wavelengths) \\
Dust Condensation     & Modified chemical   & Time-dependent nucleation \\
                      & equilibrium         & and grain growth \\
Molecular non-LTE     & Optional Extension             & Not included \\
Dust Scattering       & Isotropic                      & Isotropic \\
\end{tabular}
\label{tabCodes}
\end{table}

\begin{figure}
\begin{center}
\includegraphics[width=9.2cm]{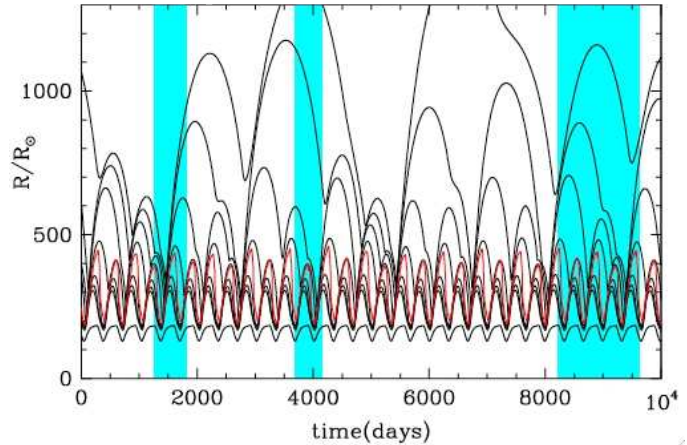}
\end{center}
\caption{An example model structure used in \citet{Ireland08} for the
  {\tt o54} series, showing
the position of the mass zones as a function of time. The layers
around the photosphere ($\sim 250$\,R$_\odot$) have a periodic motion,
while the outer layers show chaotic motions. [Electronic
version only: The red lines show the position of the optical-depth
2/3 layer and the blue shaded regions show the cycles chosen for
detailed radiative transfer modeling.]}
\label{figStructure}
\end{figure}

Qualitatively, the model structures of dynamical codes appear periodic
in the lower atmosphere ($\tau_c \sim 1$) but chaotic in the upper
atmosphere, as seen in Figure~\ref{figStructure}. Where models include
a wind, the transition to the radiatively driven wind region occurs at
$\sim$2--5 continuum-forming radii \citep[e.g Fig. 2 of
][]{Hofner03}. Even in the deepest layers, the 
motion is not sinusoidal due to the strong non-linear effects (most
importantly, shocks) that damp the pulsations.

The majority of observed properties of pulsating, extended AGB stars,
including wavelength-dependent interferometric diameters
\citep{Woodruff09}, low- and 
high- resolution spectra, light curve
shapes and phase-lags between 
wavelengths (Ireland et al, 2011, submitted) 
can all be reproduced to some extent by the modern codes in certain
instances. A few key model properties and challenges will be presented
below, focusing on the {\tt CODEX} models which the author is most
familiar with.

%{\bf Do the models {\em work}?}
%Models can reproduce light-curves, spectra and interferometry {\small
%  with a few caveats} 
%\begin{figure}
%\plotone{o54_lcurves.eps}
%\plotone{hband_codex.eps}
%\plotone{kband_codex.eps}
%\caption{COPYRIGHT: {\tt o54} series. left $o$~Ceti, right R~Cha
%(Ireland et al 2011, submitted).}
%\end{figure}

\subsection{Molecular Shells}

Very extended dust shells have been seen around AGB stars for decades
\citep[e.g.][]{Danchi94} but only in the past decade has it been clear
that molecular shells, particularly H$_2$O shells, exist around Mira
variables \citep[e.g.][]{Mennesson02}. Published models have naturally 
explained these water ``shells'' ever since they were observed
\citep{Tej03b} with a monotonic decrease in gas density with
increasing radius, coupled with a sharp increase in H$_2$O
condensation fraction as the temperature decreases with radius. 
These molecular shells may
therefore be quite different from shells produced where radiative
acceleration of dust becomes important, which can cause increases of
density with radius at the base of a shell being driven away
radiatively \citep[e.g.][]{Hofner03}.

%\subsection{Opacity Sources}
%
%\begin{itemize}
%\item In the outer atmospheres of M giants, ``easy'' continuous opacity
%sources do not play a role.
%\item Once dust forms, it is {\em decoupled} from the gas, as $\int
%  \kappa J \gg k_b T/\tau_c$, with $\tau_c$ the collision timescale.
%\end{itemize}
%\includegraphics[width=5.5cm]{ferguson.jpg}\\
%Ferguson et al (2005) Rosseland mean opacities ($\rho=10^{-15}T^3$)
%
%At temperatures roughly equal to the $\tau=1$ temperature accross the
%M~Giant spectral range, CO and N$_2$ form, binding the majority of the ``metal'' mass..
%For lower temperatures, T$\lesssim$2500\,K, many complex phase
%transitions occur in extended AGB atmospheres.
%
%\begin{itemize}
%\item In C stars (C/O $\gtrsim$1), dust forms rapidly. We'll consider M
%  (and S) stars from now on.
%\item TiO and ZrO have the highest bond energies of optically-active
%  molecules. SiO also forms, but important transitions are in the UV.
%\item VO forms next, and becomes important for late-M stars. 
%\item H$_2$O binds all the remaining oxygen. A small amount of OH also
%  forms but is never dominant.
%\item TiO$_2$ easily forms long molecule chains then nuclei, enabling
%  dust to form.
%\item As dust becomes Fe-rich, it ends up with an opacity of 5
%  (0.02/Z) cm$^2$/g, or potentially a little more if grains are large
%  with a scattering contribution.
%\end{itemize}

\subsection{Mass Loss and Radiation Pressure}

The details of the mass-loss process in extended AGB stars is the
subject of much active research and debate
\citep[e.g.][]{Woitke06,Hofner08}. Even simple questions, like 
{\em ``can pulsation alone drive measurable mass loss?''} have not been
clearly answered in the literature, as no self-excited code has the
mass resolution to resolve mass loss rates smaller than
$10^{-7}$\,M$_\odot$\,yr$^{-1}$. However, it is generally accepted
that only a combination of pulsation and radiative acceleration can
cause the significant mass loss rates of
$>10^{-7}$M$_\odot$\,yr$^{-1}$ seen in ``typical'' M-type Miras.

Neglecting scattering and assuming
full dust condensation in chemical equilibrium, when L/M (luminosity/mass
in solar units) exceeds $\sim$3,000(0.02/Z) (O-rich) or
  $\sim$300(0.02/Z) (C-rich, C/O=2), radiation can drive winds. For
C-rich stars, the relatively high condensation temperature of carbon
dust means that the material does not have to be elevated very far
above the photosphere for enough dust to condense to drive a wind.
The difficulty in modelling effective mass loss in O-rich stars 
is elevating enough material to radii of $\sim$5--10 continuum-forming
radii, where Fe-rich silicates can fully condense.

An intriguing method of driving winds was recently given in
\citet{Hofner08}, by scattering from grains of several hundred nm
radius. Both the relatively small angular sizes measured in-between
TiO absorption bands in visible/near-IR interferometry and optical interferometric
polarimetry observations demonstrate that any scattering must be
optically-thin (Section~\ref{sectNearby}). However, it is still possible 
to drive a wind with only a small optical depth in scattering, 
as long as the shell mass is small. 

The
most direct observational constraint on the capability for dust
scatting to drive winds is to simultaneously consider a tracer of gas
mass and a tracer of dust scattering optical depth. At moderate or low
spectral resolution, the shell mass is best probed with
interferometric observations in the near infrared (1.2 through to 4
microns) that measure the H$_2$O optical depth and enable the shell
mass to be inferred. 

To date there is no paper clearly combining column density constraints
of H$_2$O and dust scattering. What has been done is to compare
dynamical models of Mira variables to observations in the
near-infrared \citep{Woodruff09} and to independently compare these
models to diameter measurements at wavelengths $<1$\,$\mu$m where
scattering dominates \citep{Ireland06,Ireland08}. The models are only
able to simultaneously reproduce the extension at wavelengths
sensitive to H$_2$O and the smaller diameters in-between TiO bands at
e.g. 830\,nm if the grain radius of silicate grains is less than
approximately 70\,nm.

%\begin{figure}
%\plotone{vis_diams.eps}
%\caption{Placeholder figure showing the observed diameter versus
%  wavelength, preferably from vis to IR. Maybe a table?}
%\end{figure}

\subsection{Non-LTE in Extended Atmospheres}

\begin{figure}
%\plotone{extend_eg.eps}
\begin{center}
\includegraphics[width=10cm]{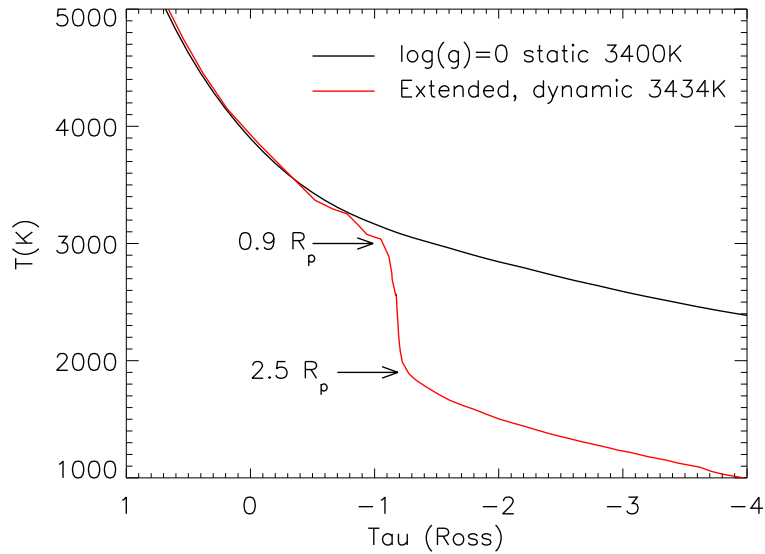}
\end{center}
\caption{The comparison of a static and dynamic LTE model atmosphere
  temperature profile with similar log$(g)$ and effective
  temperature. The dilution of the radiation field due to 
 geometrical extension causes the temperature to be very much reduced
 at small optical depths. $R_p$ is the ``parent-star'' static model
 radius.}
\label{figExtended}
\end{figure}

For the strong electronic transitions of TiO that define the M-type
spectral sequence, typical Einstein~A coefficients are
$\sim10^7$\,Hz. This is much higher than collision rates which are
$<10^3$\,Hz in the TiO forming layers for log$(g)<0$. Despite this,
static models reproduce TiO bands reasonably well
\citep[e.g.][]{Plez98}. For very extended dynamical models, however,
the Local Thermodynamic Equilibrium (LTE) assumption reproduces 
observed band depths poorly. The reason
for this can be seen in Figure~\ref{figExtended}, where the lower
temperatures in the outer layers of the dynamic model causes much
deeper than observed bands. This kind of temperature difference makes
a factor of $\sim$100 difference in the Planck function in e.g. the
deep TiO absorption feature at 670\,nm. As no features have depths of
more than a factor of a few relative to the neighboring
pseudo-continuum, non-LTE effects play a very significant role in the
$\sim$450-900\,nm region of M-type Mira spectra. A code that includes
these non-LTE effects is therefore essential if model comparisons to
visible light curves or spectral types are to be made.

%1 g cm^{-1} s^{-2} = 0.1 kg/m/s.
%Pressure of 10 dynes/cm^2 = 1 kg/m/s
%-> 3.6e19 particles/m^3
%velocities are 2800 m/s.
%Assume a 1 Angstrom collisional radius. Then
% A v n = 1000 collisions/s

%\begin{figure}
%\plotone{m7.eps}
%\end{figure}

At the time of writing, no extended code yet solves for the level populations
of molecules self-consistently. The approximation used in the {\tt
  CODEX} models is a fluorescence scattering approximation, where the
rovibrational states and the relative population in the ground singlet
and ground triplet state are in LTE, but transitions to and from the
excited electronic states are modeled as a scattering process with
appropriate wavelength redistribution. Details of this method are given
in \citet{Ireland08}. As the existing database of visible light curves
and visible spectral types is so rich, it is essential that in time
models can make reliable predictions in the visible wavelength range.

%\begin{equation}
% S(\lambda) = \frac{\kappa(\lambda) B(\lambda)+ \int \sigma(\lambda,\lambda') J(\lambda') d\lambda'}
% {\kappa(\lambda) + \int \sigma(\lambda',\lambda)d\lambda'}. \nonumber
%\end{equation}}
%\plotone{nonlte_tio.eps}

\section[Goals]{Summary and Brave Future Goals}
\label{sectFuture}
%What can we observe out to Mpc distances?
%\begin{itemize}
%\item zYJHK light curves for Miras.
%\item zYJHKLM low-res spectra (JWST, WFIRST).
%\item V-band maxima.
%\end{itemize}

It is clear that the wealth of observations already available of extended AGB
stars can not be accurately modeled. The state of observational
knowledge 33 years ago included visible light curves, the pioneering
multi-wavelength interferometry of \citet{Labeyrie77} and spectral
types in the visible. No current modeling effect to date can
reproduce these observations and give insight into the underlying
parameters of the prototypical Mira variable $o$~Ceti. However, much
progress has been made, with current models qualitatively reproducing
all the observed features and the only matter of serious debate being
the dominant factors in pulsating AGB star mass-loss and the best
parameter choices for modeling convection.

In the next decade, imaging extended atmospheres will become routine,
and with polarimetric measurements, dust shells will be resolved as a
function of pulsation phase for a number of nearby typical Miras. But
more important than these nearby details will be the contribution of
time-domain astronomy with the world's largest telescopes. 

In the Virgo cluster, individual Miras have apparent K-magnitudes of
$\sim$23 and individual Miras in the Coma cluster will have
K-magnitudes of $\sim$27. These magnitudes are within reach of JWST or
large ground-based Extremely Large Telescopes.
For the ellipticals in Virgo in particular,
these Miras could be a superb tool for extragalactic archaeology:
determining the star-formation history and indirectly the merger
history of distant galaxies. This will be possible because such a
wealth of information (broad-band colors including colors sensitive to
molecular band depths, period, amplitude) will be available for no other single star in an
old stellar population. With this information, the fundamental
parameters of composition, initial mass and evolutionary state will be
able to be derived for a great number of Miras.

Of course, none of this will be possible unless our understanding of
extended AGB star atmospheres is first tested and calibrated against
nearby Mira prototypes. We need to establish that our treatment of
convection in self-excited pulsation is adequate and have an calibrated
prescription for the free parameters, and we need to come to an
understanding and consensus on the most important parts of a dust
formation model that can drive stellar winds. In the author's opinion,
these challenges are not insurmountable, and significant progress will
be made prior to the next {\em Why Galaxies Care} conference.

\acknowledgements The Author would like to thank B.~Norris for making
the polarimetric reduction of W~Hya available in advance of
publication, and would like to thank M. Scholz for helpful
discussions. M.~Ireland is supported by an Australian Research Council
Postdoctoral Fellowship (project number DP0878674).

\bibliographystyle{asp2010}
\bibliography{ireland}

\end{document}